\documentclass{aa}     
\usepackage{graphicx}
 
\def\cc{\,{\rm cm^{-3}}}
\def\cm2{\,{\rm cm^{-2}}}
\def\kms{\,{\rm {km\,s^{-1}}}}
\def\ergs{\,{\rm {erg\,s^{-1}}}}

\def\h2{\,{\rm H_{2}}}
\def\13co{\,{\rm ^{13}CO}}

\def\degrees{\rm ^{\circ}}
\def\minutes{\rm '}
\def\seconds{\rm ''}
\def\aua{{\rm A\&A} }
\def\auas{{\rm A\&AS} }
\def\apj{{\rm ApJ} }
\def\aj{{\rm AJ} }

\def\apjl{{\rm ApJL} }
\def\araa{{\rm ARAA} }
\def\mnras{{\rm MNRAS} }

\def\pasa{{\rm PASAustral.} }

\begin{document} 
 
\title{The peculiar nebula Simeis~57}
 
\subtitle{I. Ionized Gas and Dust Extinction} 
 
\author{F.P. Israel
	\inst{1},
        M. Kloppenburg 
        \inst{1},
        P.E. Dewdney
        \inst{2}
        \and J. Bally
	\inst{3}
} 

\offprints{F.P. Israel} 
 
\institute{Sterrewacht Leiden, P.O. Box 9513, 2300 RA Leiden, the Netherlands 
\and Dominion Radio Astrophysical Observatory, Box 248, Penticton, 
     B.C., V2A 6K3, Canada
\and Department of Astrophysical and Planetary Sciences and Center for 
     Astrophysics and Space Astronomy, University of Colorado, Campus Box 389, 
     Boulder, CO 80309-0389, USA}

\date{
Received ????; accepted ????}
 
\abstract{We present high resolution radio continuum maps of the
Galactic nebula Simeis~57 (= HS~191 = G~80.3+4.7) made at the 
Westerbork Synthesis Radio Telescope and the Dominion Radio 
Astrophysical Observatory at frequencies of 609, 1412 and 1420 MHz. 
At optical and at radio wavelengths, the nebula has a peculiar ``S'' 
shape, crossed by long, thin and straight filaments. The radio maps, 
combined with other maps from existing databases, show essentially all 
radio emission from the peculiar and complex nebula to be thermal and 
optically thin. Although neither the distance nor the source of excitation 
of Simeis~57 are known, the nebula can only have a moderate electron 
density of typically $n_{\rm e} = 100 \cc$. Its mass is also low, not 
exceeding some tens of solar masses. Peak emission measures are 5000 
pc cm$^{-6}$.Obscuring dust is closely associated with the nebula, but 
seems to occur mostly in front of it. Extinctions vary from $A_{\rm V}$ = 
1.0 mag to $A_{\rm V}$ = 2.8 mag with a mean of about 2 mag. The 
extinction and the far-infrared emission at $\lambda 100\mu$m are 
well-correlated. 
\keywords{Nebulae: individual: Simeis~57 - HS 191 -- ISM: HII regions -
radio continuum} 
} 

\authorrunning{F.P. Israel et al.}
\titlerunning{Radio maps of Simeis~57}

\maketitle

\section{Introduction} 

\begin{figure}
\unitlength1cm
\begin{minipage}[b]{8.9cm}
\resizebox{10.5cm}{!}{\includegraphics*{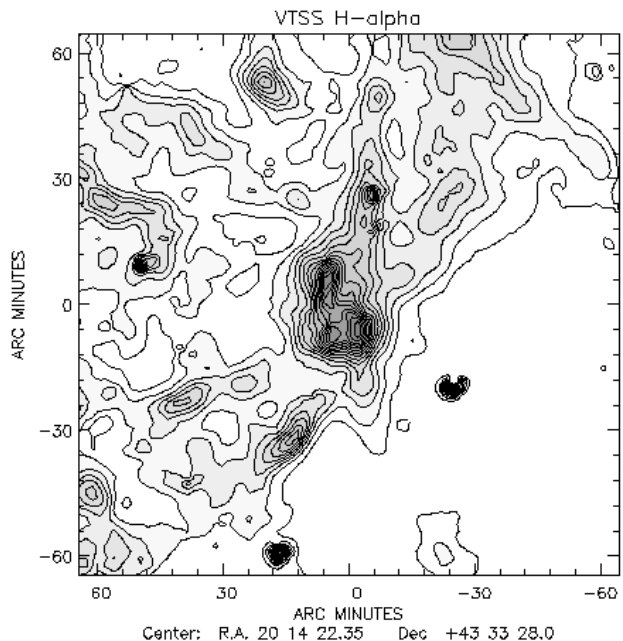}}
\end{minipage}
\caption[]{H-$\alpha$ map from the Virginia Tech Spectral-Line Survey
(http://www.phys.vt.edu/\~\rm halpha/). 
Contours are at 20, 40, 60, ... Rayleigh (1 Rayleigh = 10$^{6}/4\pi$ 
photons cm$^{-2}$ ster$^{-1}$). 
Simeis~57 is at center, and consists of the H-$\alpha$ clouds DWB 111, 
118 and 119 cataloged by Dickel et al. (1969). Other clouds are DWB 107 
(+15, -30), DWB~117 (+40, -25), DWB~120 (-20, +25), DWB~136 (+20, +55) 
and DWB~131 (+50, +20).
}
\label{vtss}
\end{figure} 

Simeis 57 (also known as HS~191 and C~191; Gaze $\&$ Shajn 1951, 1955) is
a high surface-brightness nebula in the constellation Cygnus, about two
degrees north of the large nebula IC~1318a. It is a prominent object on 
the Palomar Sky Survey images, not least because of its peculiar S-shape, 
reminiscent of a garden sprinkler. Various parts of the complex nebula 
are listed separately in the catalogue of optically visible HII regions
in the Cygnus X region by Dickel et al. (1969; hereafter DWB). The southern
arm of the "S" is included as DWB~111, the northern arm as DWB~119. 
Projected onto the center of the "S" is a faint "bow tie" of optical 
emission {\it and} absorption, consisting of a broad north-south filament 
(DWB~107, 118, 126) and a narrow, fainter northeast-southwest filament
(DWB~108 and 125, perhaps DWB~136 as well). This bow tie extends over 
roughly 1.5$\degrees$ from the supernova remnant W63 in the north towards 
the bulk of the Cygnus X region in the southeast. These features are quite 
distinct, even though the whole field of view is very confused with 
nebulosity and extinction patches belonging to the Orion (Local) Arm 
which is viewed mostly tangentially at these longitudes. This, and the 
small radial velocities $V_{\rm LSR} \approx -12 \kms$ (Dixon et al. 1981;
Piepenbrink $\&$ Wendker 1988) associated with Simeis~57 do not allow 
a kinematic distance determination, nor has the association of any star
with the object been established. Its actual distance is thus unknown, 
although both its Galactic latitude ($l=80.4\degrees$, $b=+4.7\degrees$) 
and its angular extent (over 20$'$) suggest that it is a relatively nearby 
object, as was also concluded by Dickel et al. (1969). 

Despite its remarkable appearance and its apparent vicinity, very
little attention has been paid to Simeis~57 since it was first catalogued 
in the 1950's.  The nebular complex corresponds to the radio source W~61 
(Westerhout, 1958) and also appears to be the counterpart of 3C 425. Much 
of the existing information is contained within the various Cygnus X 
region surveys (e.g. Wendker 1967, 1970; Dickel et al. 1969; Wendker et 
al. 1991; for a compilation of early maps often including HS~191, see 
Goudis 1976). Its H$\alpha$ emission was measured by Karyagina $\&$ 
Glushkov (1971) and a calibrated H$\alpha$ map was included in one of the
Cygnus~X region surveys by Dickel $\&$ Wendker (1977). A set of [NII] 
radial velocities across the nebula was published by Dixon et al. (1981). 
Simeis~57 was also included in the Virginia Tech Spectral-Line Survey
(Dennison et al. 1998). The relevant 1.6$'$ resolution H$\alpha$ image 
of the nebula and its surroundings is shown in Fig. \ref{vtss}. Note that 
in this Figure, and throughout the remainder of this paper, equatorial 
coordinates are given with respect to the 1950.0 equinox.

\section{Observations}

\subsection{WSRT 608.5 MHz and 1412 MHz continuum maps}

We observed  fields centered on Simeis~57 with the Westerbork Synthesis 
Radio Telescope (WSRT) in the radio continuum at 608.5 ($\lambda$ = 49 cm)
and  1412 MHz (($\lambda$ = 21 cm). The WSRT is an aperture synthesis 
interferometer consisting of a linear array of 14 antennas of diameter
25 m, arranged on a 2.7 km east-west line. Ten of the antennas are on 
fixed mountings, 144 m apart. The remaining four telescope are movable, 
enabling recovery of visibilities corresponding to baselines from 36 m to 
2.7 km. 

At 1412 MHz, two sets of 12-hr observations were taken on May 16-17 and 
July 23-24, 1982 respectively. At 608.5 MHz, five sets of 12-hr 
observations were taken on July 22, August 9, September 1, September 6 and
September 9, 1988 respectively. The observation on September 6 was a repeat
of an unsatisfactory result obtained on September 1, which we have not used
in the following.

All data reduction was done in $\cal AIPS$. Although we did check the
emission of Simeis~57 for signs of polarization, none was found. We
did have to address, however, the complications caused by the presence 
of the strong double radio source Cygnus A, about 4.3$\degrees$ from the 
fringe stopping center. This source, although weakened, distorted and 
spuriously polarized by the primary beam sidelobe response of the WSRT, 
still causes relatively strong grating rings to cross the observed field 
and confuse the Simeis~57 image. At both frequencies, we constructed small 
maps of the Cygnus A area, which we then cleaned. These clean-components 
were translated back into the (u,v)-plane and subtracted from the 
(u,v)-data of Simeis~57 before these were Fourier-transformed into a 
sky image. This procedure worked very well at 1412 MHz (were the
Cygnus A response is weakest). It yielded an acceptable result at 608.5 
MHz, although some of the remnant grating response can still be seen in
the western half of the sky map.

Virtually all of the emission of interest in the Simeis~57 field is extended.
Normal cleaning methods (such as the type devised by Clark and enhanced by 
Cotton $\&$ Schwab, which is the standard implemented in $\cal AIPS$) 
do not handle extended emission very well, because the point-component 
model they use is too far from the actual reality. As the Simeis~57 
WSRT fields contain both pointlike and extended sources, the clean-method 
of Cornwell $\&$ Holdaway (July 1999, Socorro imaging conference) can be 
used. This method is a multi-resolution modification of clean. The full 
resolution image is convolved with gaussians of different widths while 
a dirty beam appropriate to a component of each width is constructed. In 
this way several maps with decreasing resolution are obtained. One of 
the convolving gaussians has "zero-width" to include the point-source model. 
All these maps are simultaneously cleaned to obtain clean components with 
varying extent, thus providing a much beter response to extended structure.
$\cal AIPS$ 31DEC00 has a (u,v)-based variation of this algorithm implemented 
As the method is (u,v)-based, the translated clean components are subtracted
directly from the (u,v)-plane. This yields a better result than subtraction 
of scaled dirty beams from the sky map. 

\begin{figure*}[]
\unitlength1cm
\begin{minipage}[]{8.9cm}
\resizebox{10.5cm}{!}{\includegraphics*{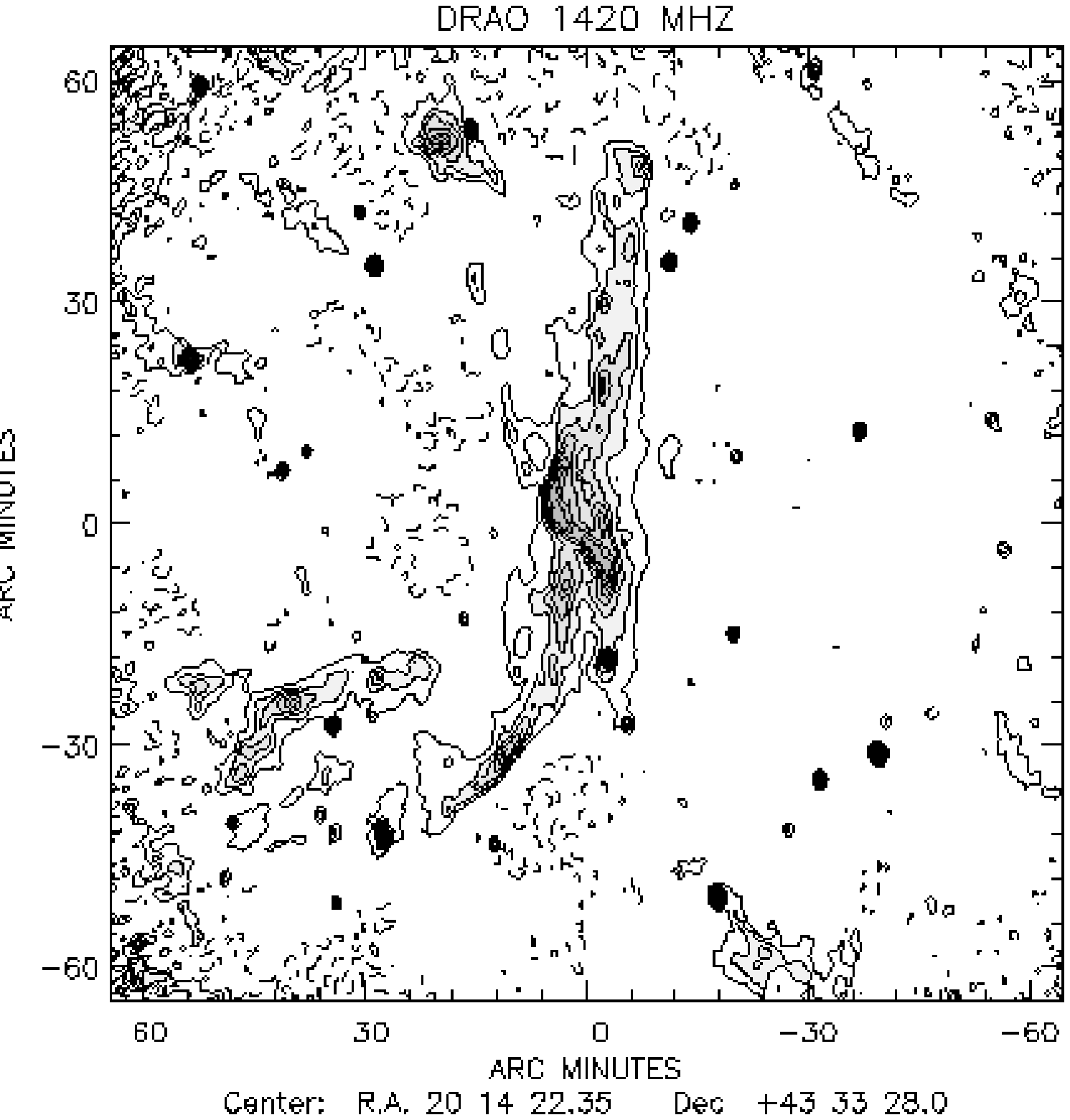}}
\end{minipage}
\begin{minipage}[]{8.9cm}
\resizebox{10.5cm}{!}{\includegraphics*{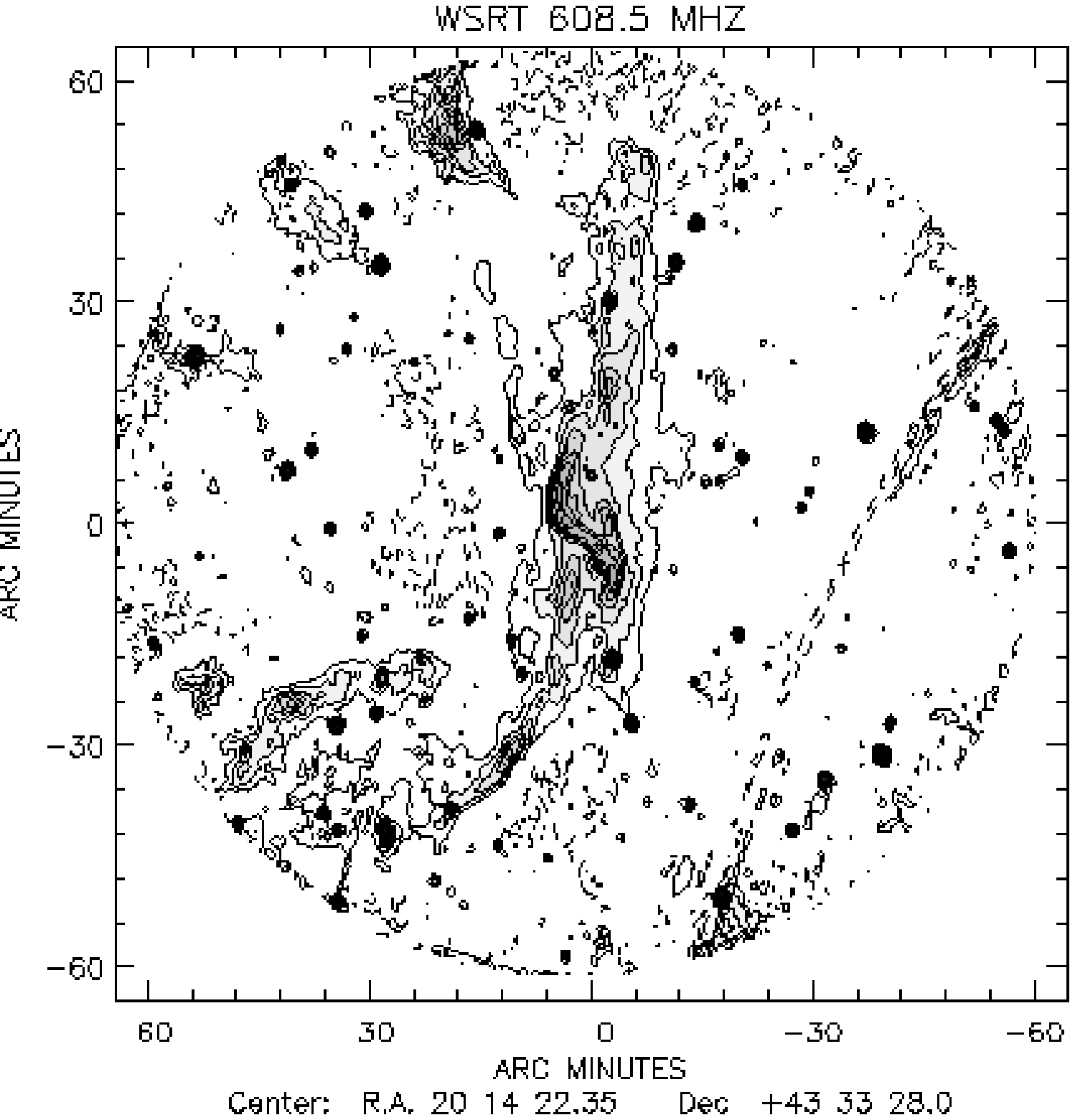}}
\end{minipage}
\caption[]{Radio continuum maps of Simeis~57 and surroundings. Left: 
DRAO 1420 MHz map. Contours are at 8 $\times$ (-2, -1, 1, 2, 3, \dots) 
mJy/beam (=$58\times80''$). Increasing noise at the map edges is due to 
increasing primary beam correction. A grating ring from the nearby, 
strong radio source Cygnus A is visible in the southwest corner of the map.
Right: WSRT 608.5 MHz map. Contours at 4 $\times$ (-2, -1, 1, 2, 3, 
\dots) mJy/beam (=$44\times60''$)
}
\label{radiolarge}
\end{figure*} 

\begin{figure*}
\unitlength1cm
\begin{minipage}[b]{18.0cm}
\resizebox{22.0cm}{!}{\hspace{-1.5cm}{\includegraphics*{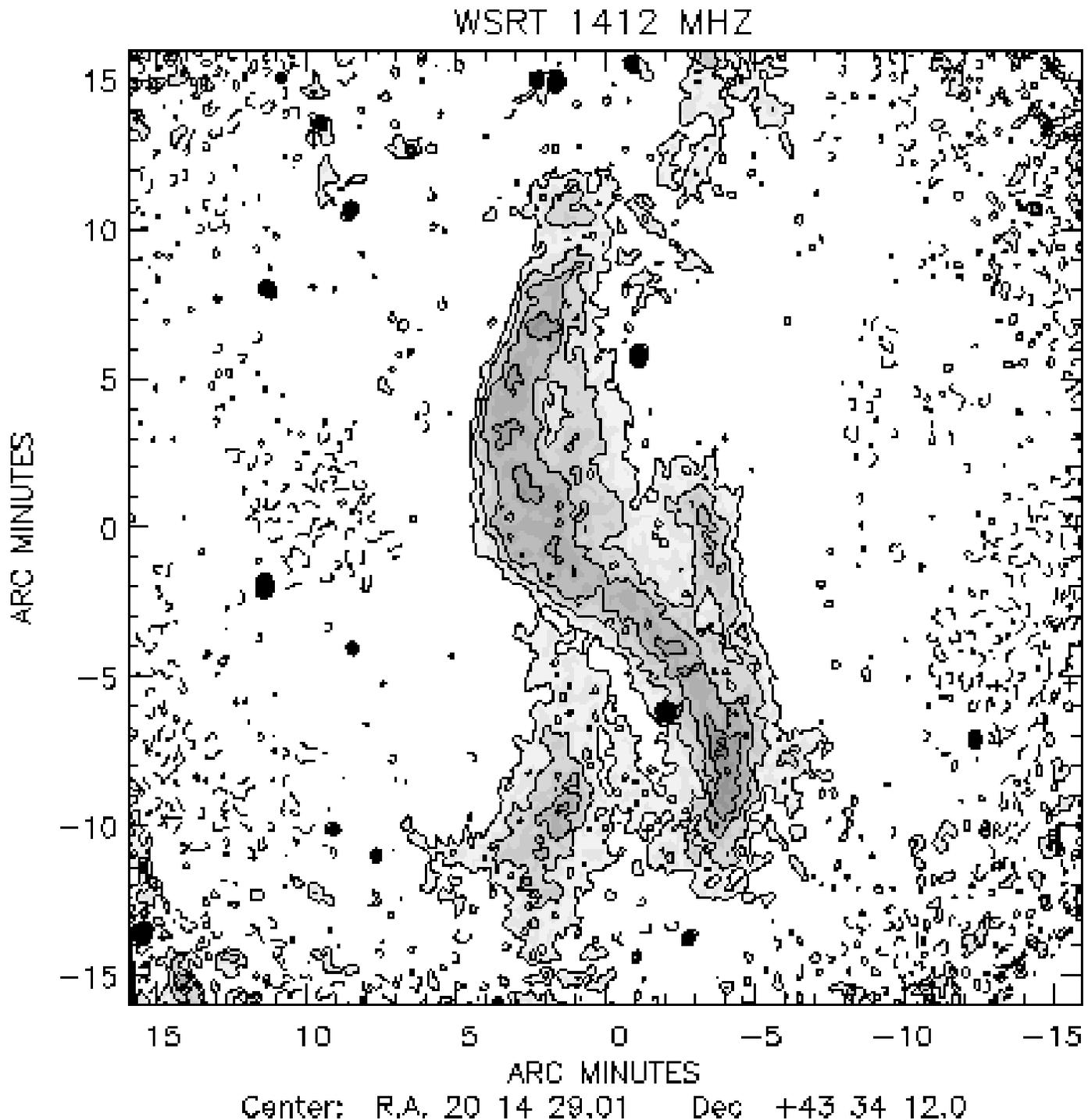}}}
\end{minipage}
\caption[]{WSRT radio continuum map of Simeis~57 at 1412 MHz. Contours 
are at 0.8 $\times$ (-2, -1, 1, 2, 3, \dots) mJy/beam (=$16\times21''$).
}
\label{radiocenter}
\end{figure*} 

\begin{table*}
\caption[]{\centerline{Observational parameters}}
\begin{center}
\begin{tabular}{lrrr}
\hline
\noalign{\smallskip}
Telescope                      & WSRT   	     & WSRT           & DRAO          \\
Frequency (MHz)                & 608.5   	     & 1412           & 1420          \\
Wavelength (cm)                & 49.3   	     & 21.2           & 21.1          \\
Observing Time (hrs)           & $4\times 12$        & $2\times 12$   & $35\times 12$ \\
Observation Date               & Jul-Sep 1988        & May/Jul 1982   & Sep 1985      \\
RA Phase Center (B1950)        & 20$^{\mathrm h}$14$^{\mathrm m}$36.0$^{\mathrm s}$  
                               & 20$^{\mathrm h}$14$^{\mathrm m}$36.0$^{\mathrm s}$
                               & 20$^{\mathrm h}$14$^{\mathrm m}$32.0$^{\mathrm s}$  \\
DEC Phase Center (B1950)       & +43$\degrees$35$\minutes$00$\seconds$ 
                               & +43$\degrees$35$\minutes$00$\seconds$ 
                               & +43$\degrees$33$\minutes$13$\seconds$ \\

RA Phase Center (J2000)        & 20$^{\mathrm h}$16$^{\mathrm m}$16.8$^{\mathrm s}$  
                               & 20$^{\mathrm h}$16$^{\mathrm m}$16.8$^{\mathrm s}$
                               & 20$^{\mathrm h}$16$^{\mathrm m}$12.9$^{\mathrm s}$  \\
DEC Phase Center (J2000)       & +43$\degrees$44$\minutes$18$\seconds$ 
                               & +43$\degrees$44$\minutes$18$\seconds$ 
                               & +43$\degrees$42$\minutes$31$\seconds$ \\

Observed Spacings (m)          & 36 - 2754          & 36 - 2736      & 12.9 - 604.3  \\
Spacing Increment (m)          & 18                 & 36             & 4.29          \\
HPBW Primary Beam ($\degrees$) & 1.45               & 0.60           & 1.64          \\ 
HPBW Synthesized Beam($''$)    & $44\times 60$      & $16\times 21$  & $58\times 80$ \\
Total Bandwidth (MHz)          & 2.5                & 10             & 15            \\
R.m.s. Noise (mJy/beam)        & 2                  & 0.4            & 0.8           \\
Map Cell Size ($''$)           & $8\times 8$        & $4\times 4$    & $30\times 30$ \\
CLEAN Limit (mJy/beam)         & 2  	            & 2              & 2             \\
No. CLEAN components           & see text           & see text       & 30 000        \\
Calibrators                    & 3C147 (38.2 Jy)    & 3C147 (22.0 Jy)  & 3C147 (22.0 Jy) \\
                               & 3C286 (20.8 Jy)    & 3C286 (14.8 Jy)  & 3C295 (22.1 Jy) \\
\noalign{\smallskip}
\hline
\end{tabular}
\end{center}
\label{obspar}
\end{table*}

At 1412 MHz we used a cell size of 4$''$ and five different gaussians
for CLEANing. We CLEANed to a level of 2 mJy/beam and obtained 51, 
29, 456, 2136 and 3318 CLEAN components for gaussians with widths of 
0, 32, 64, 128 and 256 arcsec respectively. The 1412 MHz restored
synthesized beam FWHM of the CLEANed map is $16\seconds \times 21 \seconds$.
At 608.5 MHz we used a cell size of 8$''$ and again five gaussians.
Here, we used 2884, 5273, 7568, 8571 and 10496 CLEAN components for 
gaussians of width 0, 64, 128, 256 and 512 arcsec respectively. The
608.5 MHz restored synthesized beam was $44\seconds \times 60\seconds$.\\

The final maps are shown in Figs. \ref{radiolarge} and \ref{radiocenter}.

\subsection{DRAO 408 and 1420 MHz continuum maps}

We observed Simeis~57 and its surroundings in September 1985 with 
the Synthesis Telescope at the Dominion Radio Astrophysical 
Observatory (DRAO) in Penticton. The DRAO Synthesis Telescope is 
a wide-field telescope operating simultaneously in broadband 
continuum at 408 MHz ($\lambda$ = 74 cm) and 1420 MHz ($\lambda$ = 
21 cm ), as well as in narrowband (at the time 128 channels)
neutral atomic hydrogen (HI) spectral line bands. The telescope 
consisted of four 9 m paraboloids on a 600 m long east-west line.
Two of the antennas are movable, enabling the recovery of visibilities 
from 12.9 m to 604.3 m. Broad structure in the continuum 
emission, representing visibilities corresponding to spacings 
shortwards of 12.9 m was obtained from existing low-resolution 
surveys at the two frequencies. Extended structure in HI was measured 
with the 26 m antenna at DRAO. A more detailed description of the 
telescope in the form used for the observations presented here was given
by Higgs (1989), whereas its current capacities have been described by 
Landecker et al. (2000). 

The 408 MHz observations and the resulting map have already been published 
before. For a description of the observations, reduction procedures, map 
and results we therefore refer to Higgs et al. (1991). The 1420 MHz 
continuum was observed over a bandwidth of 15 MHz. We mapped a field 
of view with a diameter of 2.6$\degrees$ (i.e to 20$\%$ primary beam 
response) on a $0.5\minutes \times 0.5\minutes$ grid. The map was cleaned 
and restored with an artifial gaussian beam of (FWHM) $58\seconds \times 
80\seconds$ (cf. Landecker et al. 2000). At 408 MHz, the disturbing 
influence of Cygnus A at about 4.5$\degrees$ from the fringe stopping 
center is quite pronounced at the relatively low intensity levels 
corresponding to the emission from Simeis~57. However, at 1420 MHz, 
the grating rings from Cygnus A could be removed reasonably well from 
most of the map, although the remnant of such a grating ring remains 
visible in the southwest corner of the map. The measured map noise is 
close to the reported r.m.s. value for DRAO 1420 MHz continuum observations 
of 0.8 mJy/beam at the map center (Higgs 1989). Finally, we have corrected 
the map for the primary beam response (Fig.\ref{radiolarge}). An earlier 
version of this map, along with a similar map derived from VLA observations, 
was published before by Cornwell (1988) in order to illustrate aspects
of Maximum Entropy image restoration.

\begin{figure}
\unitlength1cm
\begin{minipage}[b]{9.0cm}
\resizebox{10.5cm}{!}{\includegraphics*{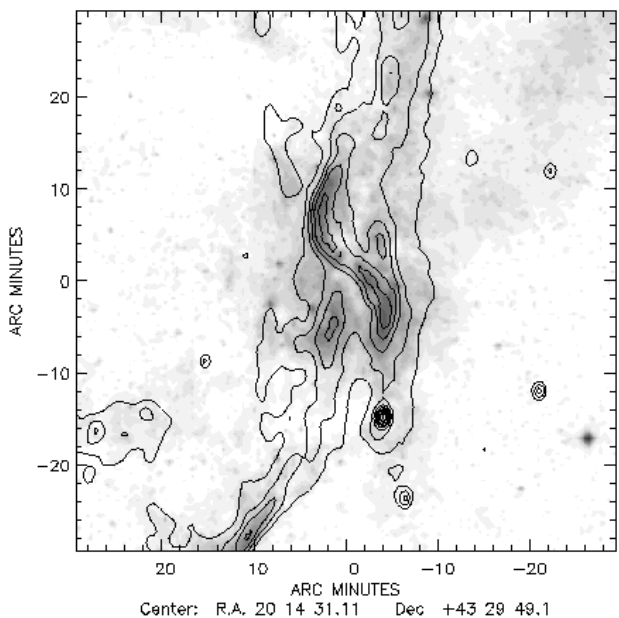}}
\end{minipage}
\caption[]{Simeis~57 21cm radio continuum emission measured with the DRAO
$58\times80''$ beam (contours at 10, 20, 30, ... mJy/beam) compared to 
first generation red Palomar Sky Survey image. Note similarity of radio 
image to optical emission {\it and} absorption. 
}
\label{PSS_drao}
\end{figure} 

A comparison of the H$\alpha$ image in Fig. \ref{vtss} with the radio 
continuum images in Fig. \ref{radiolarge} shows a good resemblance 
between the optical and radio emission. This is further illustrated in 
Fig. \ref{PSS_drao}, where we have superposed contours of 1420 MHz radio 
emission over the red PSS optical image of the nebula. The radio contours 
closely follow the outline of nebular emission, both for the ``S'' shaped 
nebula and for the north-south filament. Also note that the filament 
contains an absorption band which is most clearly seen where it crosses
the ``S''.

\subsection{Other radio continuum maps}

\begin{figure*}
\unitlength1cm
\begin{minipage}[b]{18.0cm}
\resizebox{18.0cm}{!}{\includegraphics*{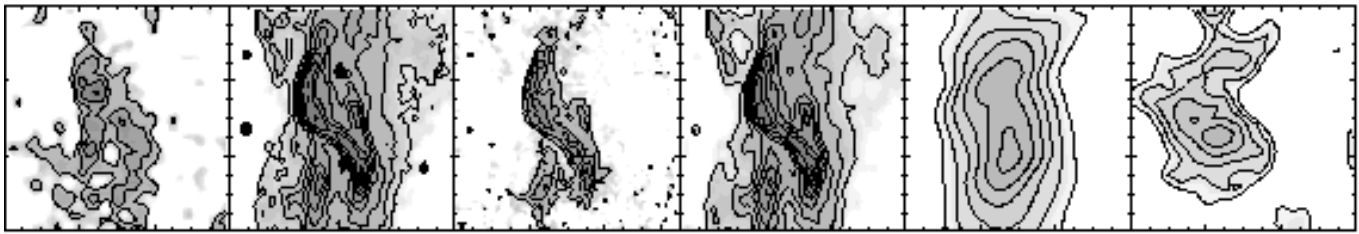}}
\end{minipage}
\caption[]{Simeis~57 at different radio continuum frequencies. From 
left to right: 327 MHz (WENSS, contours at 25 $\times$ (1, 2, 3, \dots) 
mJy/100$''$ beam), 608.5 MHz (WSRT, contours as in Fig.\ref{radiolarge}), 
1412 MHz (WSRT, contours as in Fig.\ref{radiocenter}), 1420 MHz (DRAO,
 contours as in Fig.\ref{radiolarge}), 8.35 GHz (GPA, contours at 0.12 
$\times$ (-2, -1, 1, 2, 3, \dots) Jy/670$''$beam) and 14.35 GHz (GPA, 
contours at 0.10 $\times$ (-2, -1, 1, 2, 3, \dots) Jy/480$''$beam).
}
\label{multimap}
\end{figure*}

The region containing Simeis~57 has been covered by various relatively
recent radio surveys. The first of these is the Westerbork Northern Sky 
Survey (WENSS) which includes the region at 327 MHz ($\lambda$ = 92cm)
with a resolution of $54\seconds \times 74\seconds$, a formal r.m.s. map 
noise of 3.6 mJy/beam and a limiting flux-density of about 5 $\sigma$ = 
18 mJy/beam (Rengelink et al. 1997; see also 
{\it http://www.strw.leidenuniv.nl/\verb1~1dpf/wenss/}). We convolved the
map obtained from this survey to a 100$\seconds$ circular beam in an
attempt to improve the relatively low signal-to-noise ratio (Fig. 
\ref{multimap}). However, extended emission is hardly discernible and 
even the source regions of higher surface brightness are poorly 
represented.
 
Langston et al. (2000) used the NRAO/NASA Green Bank Earth Station to 
survey the Galactic Plane in the radio continuum simultaneouisly at 
frequencies of 8.35 GHz and 14.35 GHz (see also {\it 
http://www.gb.nrao.edu/\verb1~1glangsto/GPA/}). The Galactic Plane Survey 
(GPA) covers the region ($-5\degrees < {\it b} < +5\degrees$, $-15\degrees 
< {\it l} < +255 \degrees$). The Cygnus X region containing Simeis~57 
has been mapped at both frequencies with resolutions of 670$\seconds$ and 
480$\seconds$ respectively. At the two frequencies, intensity scales are 
claimed to be better than 10$\%$ and 20$\%$ respectively. 
We determined the variance ($\sigma^2$) of the maps and used the values 
thus obtained ($\sigma$ = 0.06 Jy/beam and $\sigma$ = 0.05 Jy/beam 
respectively) as the r.m.s. noise values. The peak signal-to-noise in 
the 8.35 GHz-map is then 26, but in the 14.35 GHz map only 14. The maps 
are shown in Fig. \ref{multimap}. The 8.35 GHz map shows a clear source 
representing the nebula, even though little structure is seen, but the 
14.35 GHz map is less clear. 

Finally, the image from the NRAO VLA Sky Survey (NVSS) should be
mentioned for completeness sake. This is part of a radio continuum survey 
covering the sky north of declination -40$\degrees$ at a frequency of 
1400 MHz. Although the main features of the nebula are represented in 
this map, it is much inferior to the WSRT and DRAO maps at the same 
frequency. Simeis~57 is outside the range of the other VLA-survey, FIRST.

\section{Results and analysis}

\subsection{Spectral index}

\begin{figure*}
\unitlength1cm
\begin{minipage}[]{9.0cm}
\resizebox{8.9cm}{!}{\includegraphics*{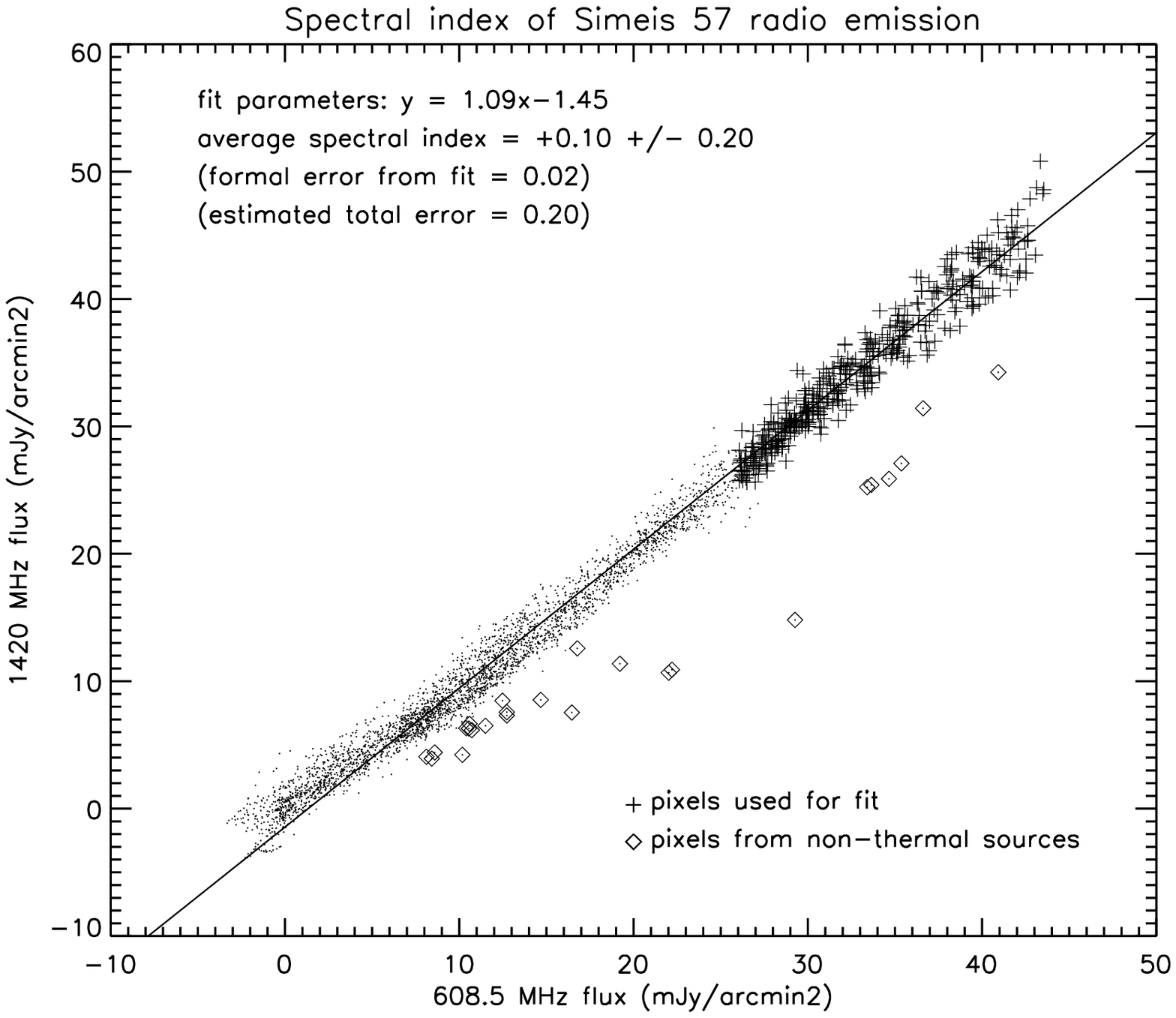}}
\end{minipage}
\begin{minipage}[]{9.0cm}
\resizebox{8.8cm}{!}{\hspace{-1.6cm}{\includegraphics*{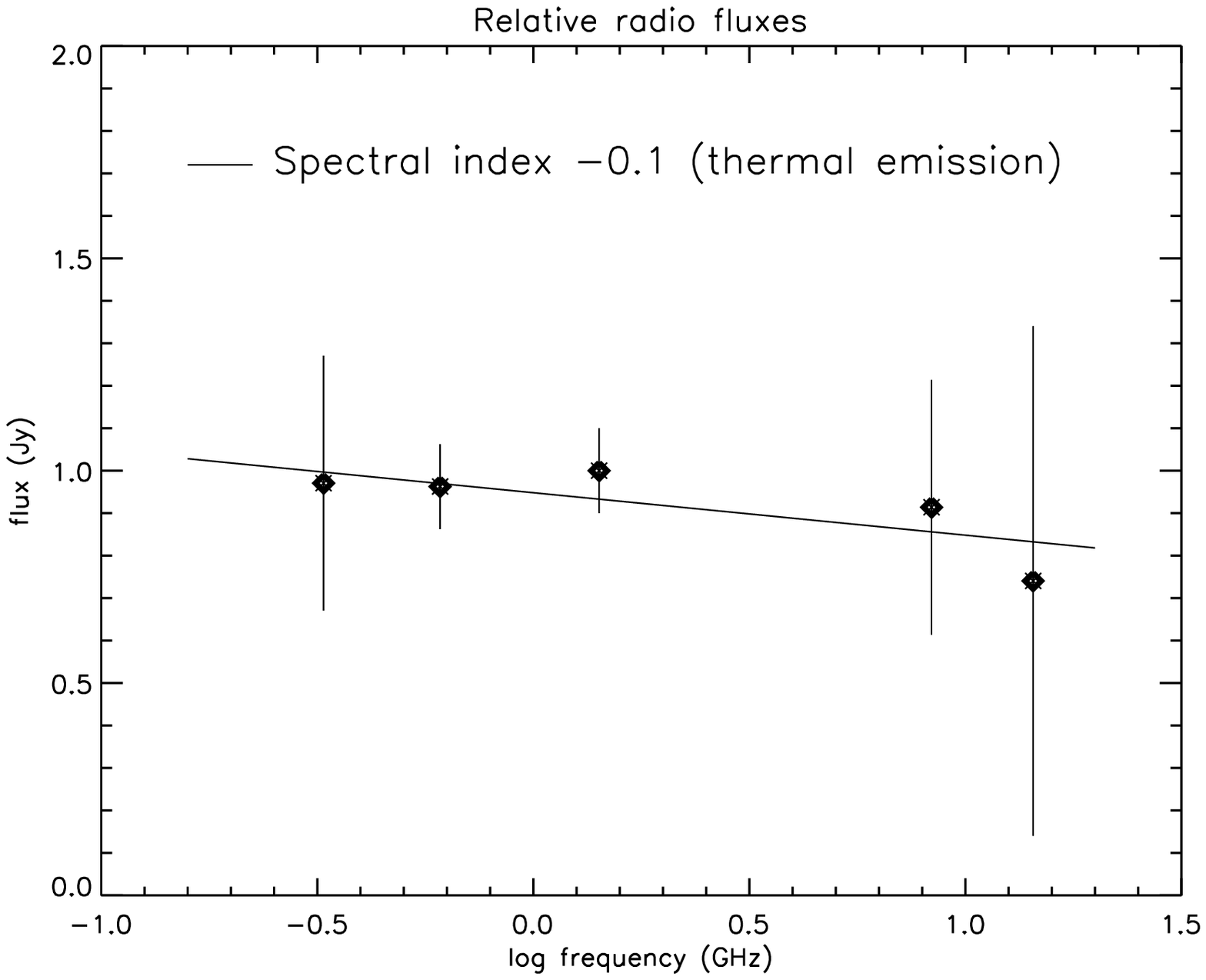}}}
\end{minipage}
\caption[]{Left: Pixel-pixel plot of WSRT 608.5 MHz map vs DRAO 1420 MHz 
map intensities. Right: Simeis~57 integrated radio flux-densities 
normalized to that at 1420 MHz, determined by pixel-to-pixel comparison. }
\label{spectralindex}
\end{figure*} 

Simeis~57 has been included in various single-dish radio surveys of 
the Cygnus X region (cf. Goudis 1976). In particular, Wendker (1967, 
1970) has presented flux-densities for Simeis~57 at various frequencies 
ranging from 610 to 4930 MHz. He found an overall spectral index 
$\alpha = +0.4$ (spectral index defined as $S_{\nu} \propto \nu^{\alpha}$) 
indicative of partially optically thick thermal emission. In his 
discussion of the object, Wendker (1967) suggested that emission from 
the north of Simeis~57 is optically thin, but that the southern part 
and especially the center are optically thick at the lower observed 
frequencies.  

We can significantly improve upon this result, because the radio data 
discussed in this paper have much higher resolution and sensitivity
than those available to Wendker. Among other things, this allows us
to separate the source from its complex background much more
accurately than was possible with the older large-beam data.

In the preceding, we have introduced radio observations of the nebula 
Simeis~57  at five different frequencies: 0.327, 0.609, 1.41/1.42, 
8.35 and 14.35 GHz (Fig. \ref{multimap}). The WSRT 1412 MHz map has the
best resolution, but the absence of low spatial frequencies in this map 
makes it unsuitable for spectral index determinations. This is, fortunately, 
not the case for the DRAO 1420 MHz and WSRT 608.5 MHz maps. The former
contains all spatial frequencies, single-dish data having been added into
the map. The latter lacks information at the shortest spacings, but its 
longer observing wavelength, its extensive clean-and-restoration, as
well as the actual angular size of the source structure of interest 
combine to reduce the adverse effects to almost complete insignificance. 
The two maps have similar resolutions and field sizes. The poorer quality 
of the WENSS 327 MHz, GPA 8.35 and GPA 14.35 GHz maps makes them less 
suitable for accurate spectral index determinations,although they can and
will be used for consistency checks.

We convolved the WSRT 609 MHz map to the DRAO 1420 MHz beam, 
and corrected both maps for small systematic base-level offsets of 
-3.4 mJy/arcmin$^2$ (DRAO) and -3.3 mJy/arcmin$^2$ (WSRT) found by 
comparing the low-intensity areas in both maps. We then constructed 
a pixel-to-pixel plot of the full corrected data set (Fig. 
\ref{spectralindex}). The source spectral index was found
by fitting all data points excluding those pertaining to the general
background and non-thermal point sources. The result obtained is
a mean spectral index $\alpha = + 0.10 \pm 0.20$, consistent with
an overall optically thin emission spectrum for Simeis~57. As the 
relatively small dispersion of points in Fig. \ref{spectralindex}
shows, this result applies not only to the bighter parts of Simeis~57, 
but also to the extended, lower surface-brightness parts of the source.

\begin{figure}
\unitlength1cm
\begin{minipage}[b]{9.0cm}
\resizebox{8.9cm}{!}{\includegraphics*{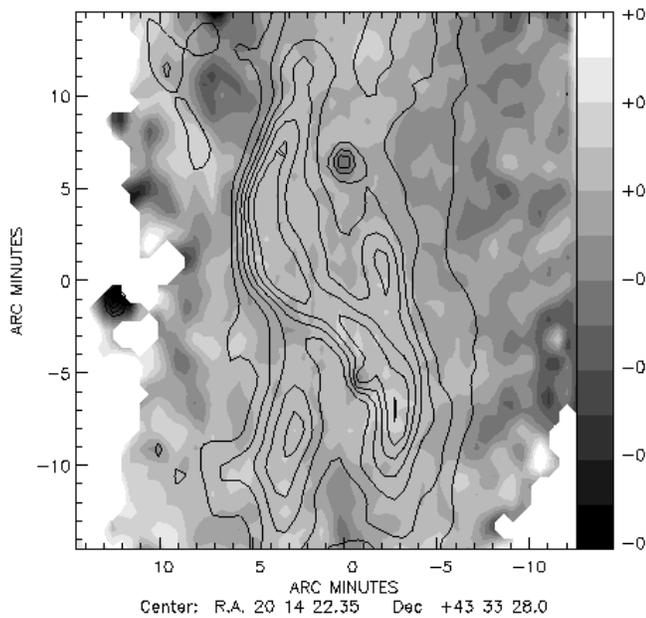}}
\end{minipage}
\caption[]{Right: Spectral index map of Simeis~57. Grayscales: map 
of the spectral index distribution. Contours at 10, 15, 20, \dots 
mJy/arcmin$^2$ outline 608.5 MHz emission
}
\label{spectralmap}
\end{figure} 

This is further illustrated by a map of the spectral index distribution, 
based on the corrected 1420 MHz and 608.5 MHz maps (Fig. \ref{spectralmap}). 
In constructing this map we have suppressed all pixels which have, at either
frequency, an intensity less than 2 mJy/arcmin$^2$ in order to prevent 
noise blow-up. The thermal nature of essentially all of Simeis~57 is again 
obvious. The only nonthermal emission in the map corresponds to point 
sources, which probably represent background radio sources unrelated to 
Simeis 57. 

We have convolved the DRAO 1420 MHz map to the lower-resolution WENSS 
327 MHz, GPA 8.35 and GPA 14.35 GHz maps, and determined average
source flux-density ratios in the same way, allowing us to derive
the relative radio spectrum shown in Fig. \ref{spectralindex}. Here, we have 
also marked the best-fitting optically thin thermal emission spectrum. 
Especially when one takes into account that the 327 MHz and 609 MHz 
flux-densities may be somewhat underestimated because of missing
spacing information, it is obvious that the emission from Simeis~57 is 
optically thin over the whole observed frequency range 0.3 -- 14 GHz. 

\subsection{Properties of the ionized gas}

As the nature of the emission is known (thermal and optically thin),
we may derive some properties of the ionized gas in which it arises.
For this purpose, we use again the DRAO 1420 MHz and WSRT 608.5 MHz maps.

\begin{figure*}[]
\unitlength1cm
\begin{minipage}[]{6.0cm}
\resizebox{5.7cm}{!}{\includegraphics*{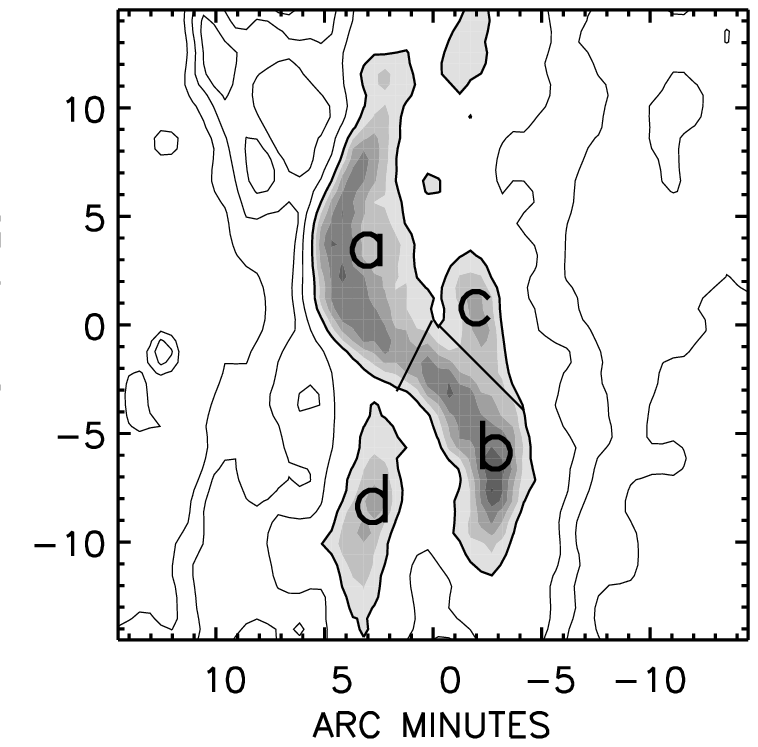}}
\end{minipage}
\begin{minipage}[]{6.0cm}
\resizebox{5.7cm}{!}{\includegraphics*{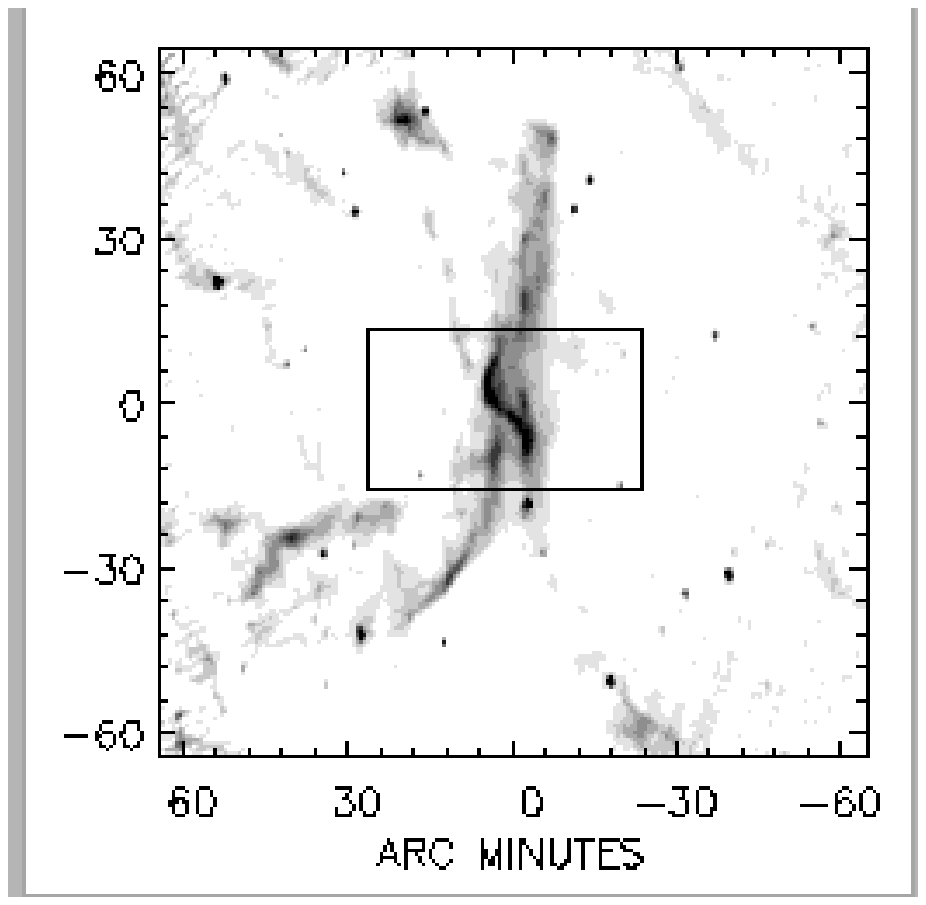}}
\end{minipage}
\begin{minipage}[]{6.0cm}
\resizebox{6.2cm}{!}{\hspace{-0.8cm}{\includegraphics*{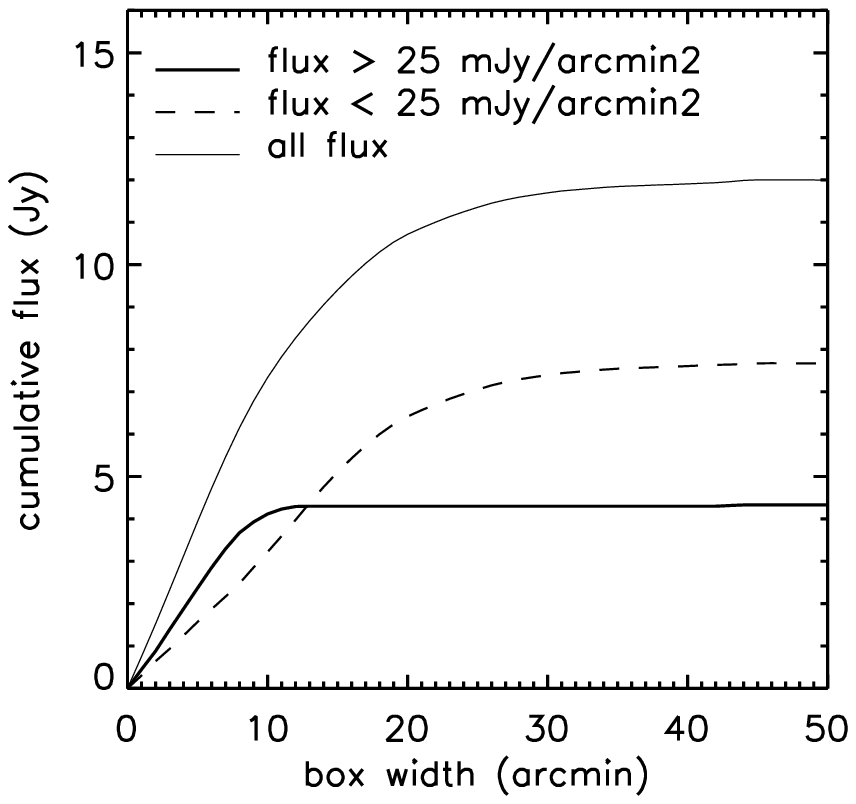}}}
\end{minipage}
\caption[]{Left: Definition of integration subregions in the DRAO 1420 MHz 
map. Contours are at 5, 10, 15, 20 and 25 mJy/arcmin$^2$. The thick contour
marks the 25 mJy/arcmin$^2$ contour used to define the four subregions.
Center: Box marking the maximum map area used in determining source total 
flux-density at 1420 MHz. Right: Total flux-density at 1420 MHz as a 
function of increasing box width}
\label{boxes}
\end{figure*} 

We have defined four different subregions making up the bright part of
the radio nebula. These regions are identified in Fig. \ref{boxes}. 
Subregions a and b (DWB~119 and 111) together form the ``S''. Subregion c 
is an extension of this and subregion d (DWB~118) is the brightest part 
of the long north-south filament.

\begin{table*}
\caption{\centerline{Observed parameters of the ionized gas}}
\begin{center}
\begin{tabular}{lccccccc}
\hline
\noalign{\smallskip}
Region  & \multicolumn{2}{c}{Dimensions} & Area      & \multicolumn{2}{c}{Flux-Density} & \multicolumn{2}{c}{Peak intensity} \\
        & Angular       & Linear         & $d\Omega$ & $S_{1420}$      & $S_{609}$      & $\sigma_{1420}$ & $\sigma_{609}$ \\
          & $(')$       & (D/500 pc)     & $('^{2})$ &  \multicolumn{2}{c}{(Jy)}        & \multicolumn{2}{c}{(Jy/$'^{2})$}\\
\noalign{\smallskip}
\hline
\noalign{\smallskip}
a=DWB~119 & 13$\times$4  & 1.9$\times$0.6 & 53       & 1.78$\pm$0.08   & 1.78 $\pm$ 0.08 & 47  & 44 \\
b=DWB~111 & 10$\times$4  & 1.5$\times$0.6 & 37       & 1.33$\pm$0.08   & 1.19 $\pm$ 0.07 & 51  & 45 \\
a+b       & 23$\times$4  & 3.4$\times$0.6 & 90       & 3.12$\pm$0.12   & 2.98 $\pm$ 0.11 & 51  & 45 \\
c         &  5$\times$3  & 0.7$\times$0.4 & 12       & 0.38$\pm$0.02   & 0.37 $\pm$ 0.02 & 39  & 37 \\
d=DWB~118 &  8$\times$3  & 1.2$\times$0.4 & 20       & 0.62$\pm$0.03   & 0.56 $\pm$ 0.03 & 39  & 37 \\
Extended  & 30$\times$20 & 4.5$\times$3.0 & 550      & 7.9             & ---             & 20  & ---\\
\noalign{\smallskip}
\hline
\end{tabular}
\end{center}
\label{radiobs}
\end{table*}

Sizes and intensities are given in Table \ref{radiobs}; note that the 
linear size is scaled to an {\it assumed} distance $D$ = 500 pc. We 
have also determined the total flux-density inside a series of boxes 
centered on the nebula, with a fixed size in declination of 30$'$ (i.e. 
corresponding to the full extent of the ``S''  nebula) and a size in
right ascension varying from 0$'$ to 50$'$. The total flux-density
measured is plotted as a function of box width $\Delta$RA (Fig 
\ref{boxes}). Essentially all flux is contained with a box with
dimensions $\Delta$RA$\times \Delta$Dec = $20'\times 30'$; the total
is $S_{1420} = 12$ Jy, to which total the bright parts mentioned above
contribute $34\%$. For the whole complex, including the north-south and
northwest-southeast filaments, Wendker (1970) found a total flux-density
$S_{2695}$ = 17.5 Jy, so that the filaments outside the maximum box 
depicted in Fig. \ref{boxes} contribute about half again of the total
flux-density inside the box.

Fig. \ref{boxes} also shows that the cumulative flux-density is constant at 
larger box widths; this is excellent confirmation that our corrections 
to the map baselevel were accurate. 

\begin{figure*}
\unitlength1cm
\begin{minipage}[]{18.0cm}
\resizebox{17.5cm}{!}{\includegraphics*{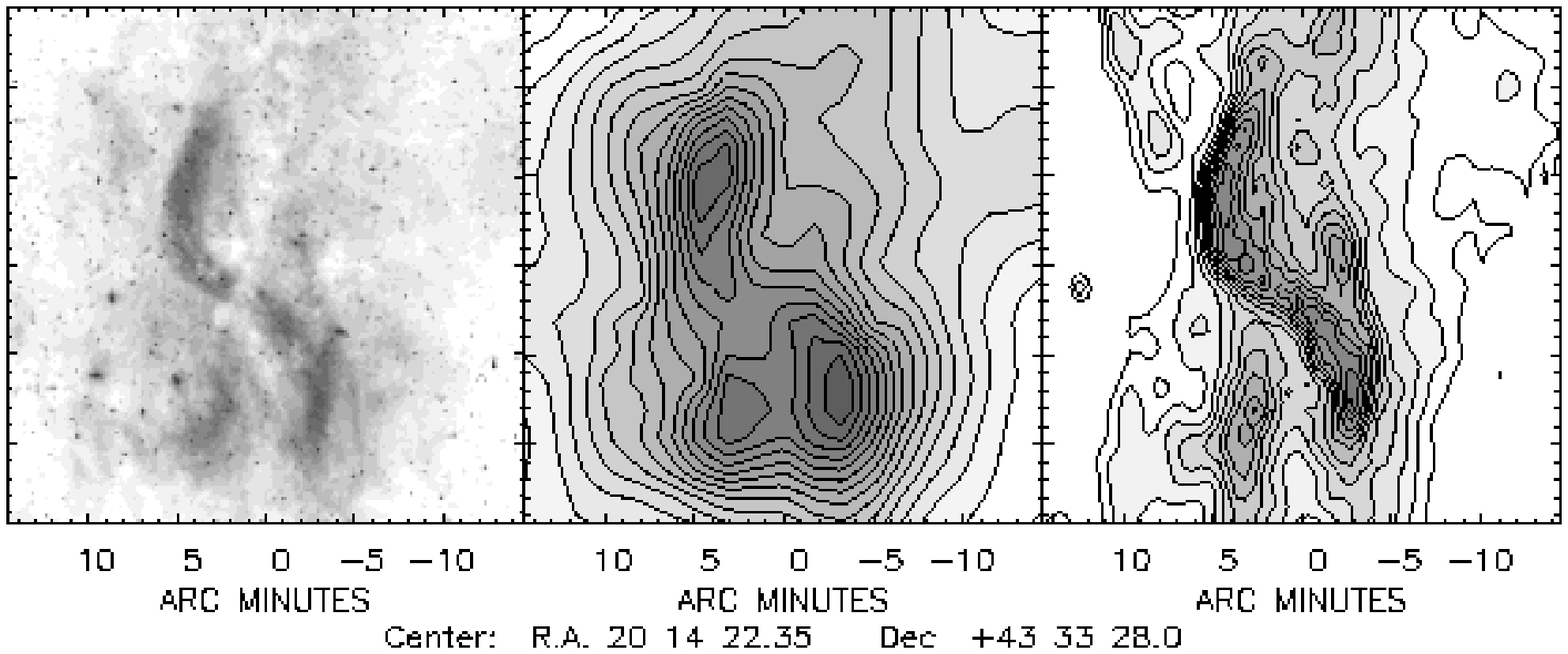}}
\end{minipage}
\caption{Comparison of Simeis 57 optical and radio images. From left to 
right: red Palomar Sky Survey, VTSS H$\alpha$ survey, and DRAO 1420 
MHz radio continuum. H$\alpha$ contours are at multiples of 20 Rayleigh.}
\label{radioptical}
\end{figure*}

\begin{table*}
\caption{\centerline{Physical properties of the ionized gas}}
\begin{center}
\begin{tabular}{ccccc}
\hline
\noalign{\smallskip}
Region   & Emission Measure & Model        & R.m.s. Electron           & Mass \\ 
         & $E.M.$           & Depth $d$    & Density $<n_e^{2}>^{1/2}$ & $M(HII)$  \\
 & $(10^{3}$ pc cm$^{-6})$ & ($D$/500 pc) & ($(D/500)^{-0.5}\cc)$     & ($(D/500)^{-2.5}$ M$_{\odot}$)  \\
\noalign{\smallskip}
\hline
\noalign{\smallskip}
a        & 5.1$\pm$0.3 & 0.6 &  93$\pm$15 &  1.2$\pm$0.2 \\
b        & 5.2$\pm$0.3 & 0.6 &  94$\pm$17 &  0.8$\pm$0.2 \\
a+b      & 5.1$\pm$0.3 & 0.6 &  94$\pm$14 &  2.1$\pm$0.2 \\
c        & 4.5$\pm$0.3 & 0.4 & 102$\pm$19 &  0.2$\pm$0.1 \\
d        & 4.4$\pm$0.2 & 0.4 & 101$\pm$20 &  0.4$\pm$0.1 \\
Extended & 2.2$\pm$0.4 & 2.9 &  28$\pm$5  &   18$\pm$3   \\
\noalign{\smallskip}
\hline
\end{tabular}
\end{center}
\label{radiophys}
\end{table*}

In deriving the physical properties in Table \ref{radiophys}, we have 
followed Mezger $\&$ Henderson (1967) for optically thin thermal
emission. We have assumed that the emitting volume is that of a 
homogeneously filled cylinder at a constant electron temperature 
$T_{e} = 10^{4}$ K, with a surface area $d\Omega$ and a depth $d$ 
corresponding to the smallest of the two projected dimensions. 
All parameters are based on an {\it assumed} distance of 500 pc; 
scaling factors for other distances are given in the header. 
We note that the ionized hydrogen masses given in Table
\ref{radiophys} are strictly upper limits because of the assumption
of homogeneity. If the gas is clumped, actual electron densities
will exceed $<n_{e}^{2}>^{1/2}$, and the mass will be accordingly
less. Nevertheless, our results indicate that electron densities 
in the extended emission region are by a factor of four substantially 
lower than in the brighter nebular parts. However, even there, actual
densities are modest and characteristic of well-evolved
HII regions (cf. Habing $\&$ Israel 1979). In contrast, most of
the mass is in the extended low-density component. The excitation
parameter of the nebula is given by $u_{n} = 13.5\, S_{1420}^{1/3}$ 
($D$/500pc)$^{2/3}$ pc cm$^{-2}$ with $S_{1420}$ in Jy. This is
related to the {\it stellar} excitation parameter by 
$u_{*} = (\Omega/4\pi)^{-1/3}\, u_{n}$, in which $\Omega$ is the 
solid angle subtended by the nebula as seen from the source of 
excitation. As the exciting star, and thus its location, is still 
completely unknown, the subtended solid angle $\Omega$ is very 
uncertain. Unrealistically assuming $\Omega = 4 \pi$, $u_{*} =
u_{n}$ = 19.5 pc cm$^{-2}$, corresponding to an exciting star
of spectral B0. More realistic values of $\Omega$ indicate 
excitation by an O5 - O8 star. At $D$ = 500 pc ($m-M$ = 8.5 mag),
a B0 star should have an {\it unreddened} magnitude $V_{o} = 5^{m}$, 
and O stars would have $V_{o} = 3^{m} - 4.5^{m}$. There are no 
obvious candidates in the vicinity of Simeis~57.

\subsection{H$\alpha$ emission and visual extinction}

\begin{figure*}[]
\begin{center}
\unitlength1cm
\begin{minipage}[]{11.9cm}
\resizebox{12.0cm}{!}{\includegraphics*{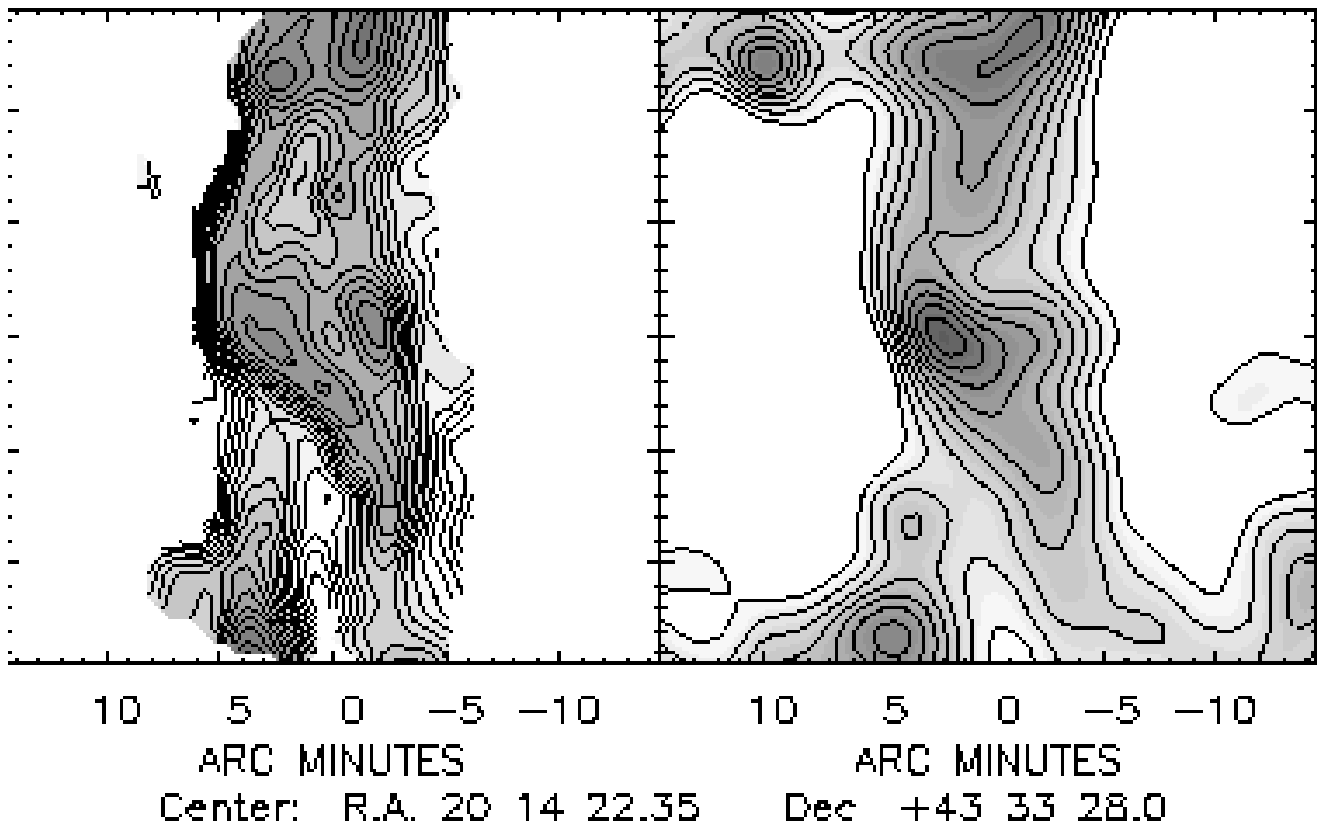}}
\end{minipage}
\begin{minipage}[]{5.8cm}
\vspace{-1.0cm}
\resizebox{6.0cm}{5.0cm}{\includegraphics*{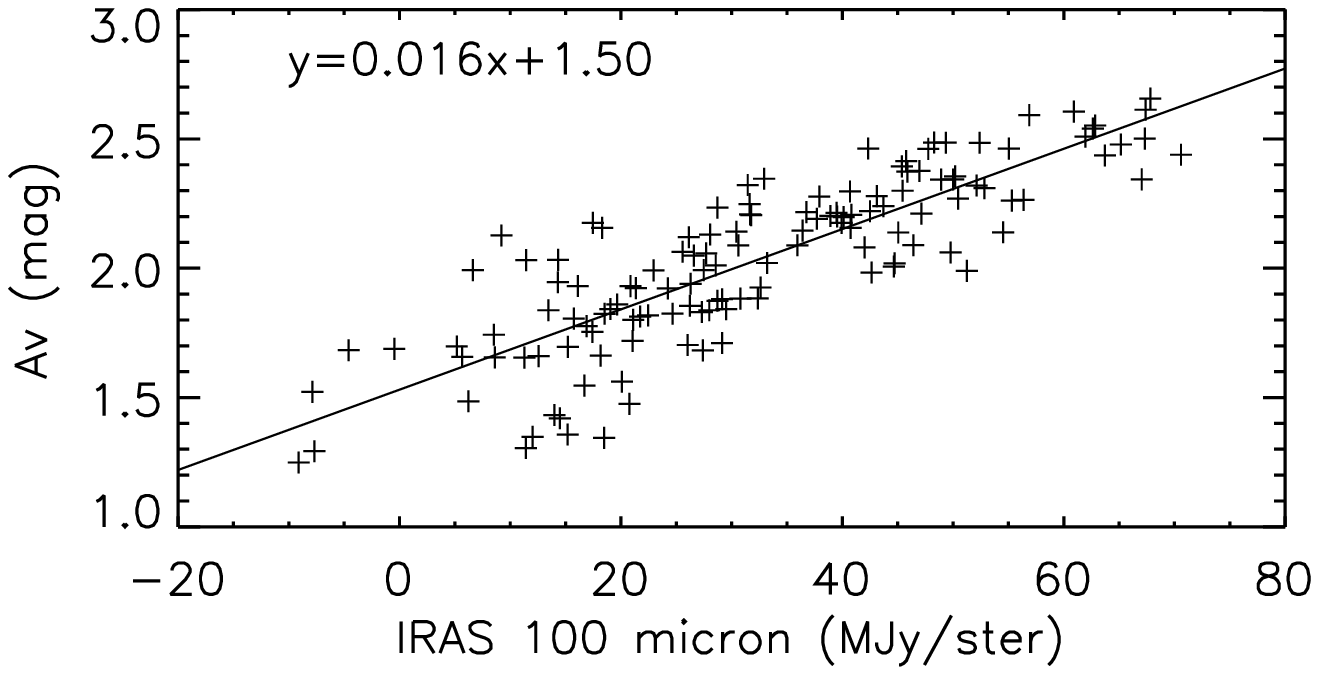}}
\end{minipage}
\caption{
Left: Map of visual extinction with contours at $A_V$ = 1.0, 1.1, 
1.2, \dots magnitudes. 
Center: IRAS/Hires map at 100$\mu$m with contours at multiples of 
10 MJy/ster.
Right: relation between visual extinction $A_{V}$ and IRAS 100
$\mu$m flux-dendsities.}
\end{center}
\label{extinction}
\end{figure*}

In Fig. \ref{radioptical} we compare, in detail, the central parts of
the optical and radio images of Simeis~57. The close similiarity is not
surprising. On the red PSS image, most of the nebular emission is expected 
to be a combination of H$\alpha$  {$\lambda$656.3 nm) and [NII] 
{$\lambda$654.8 and $\lambda$658.3 nm) line emission, i.e. similar in
origin to the optically thin free free radio continuum emission. As the 
radio continuum and optical line emission thus trace the same ionized
plasma, their distribution and intensity should be closely related.
The PSS image, although fine in its detail, is not as well-suited to a 
quantative analysis as the 1.6$'$ resolution H$\alpha$ image from the 
Virginia Tech Spectral-Line Survey (VTSS). We therefore decided to use 
the latter together with the 1420 MHz DRAO map convolved to the same
resolution.

For an assumed electron temperature $T=10^4$ K, we expect at $\nu$ =
1420 MHz an optically thin emissivity:

\begin{equation}
\epsilon_\nu = 3.9 \times 10^{-39} \; n_e^2 \;\; \mathrm{\ergs} \; \cc \; \mathrm{Hz}^{-1}
\end{equation}

\noindent
whereas the Case B H$\alpha$ emissivity (cf. Pengelly 1964) is:

\begin{equation}
\epsilon(\mathrm{H}\alpha) = 3.6 \times 10^{-25} \; n_e^2 \;\; \mathrm{\ergs} \; \cc
\end{equation}

\noindent
so that:

\begin{equation}
{{S_{1420}} \over {F_{\mathrm{H}\alpha}}} = 1.1 \times 10^{-14} \; \mathrm{Hz}^{-1}
\end{equation}

\noindent
As the H$\alpha$ emission suffers from extinction and the radio continuum
emission does not, we can use eqn. (3) to derive distribution of foreground 
extinction $A_{\alpha}$ over the entire image of Simeis 57. Assuming a dust 
extinction law $A(\lambda) \propto \lambda^{-1}$, this translates into
visual extinctions $A_{\rm V} \sim 1.2 \: A_{\alpha}$. 

The distribution of foreground visual extinction thus determined is shown
shown in Fig. \ref{extinction}. We have calculated its value only where
both $F_{\mathrm{H}\alpha}$ exceeds 100 R (1 Rayleigh $ = 10^6/4\pi$ photons 
cm$^{-2}$ s$^{-1}$ ster$^{-1}$ = $2.04 \times 10^{12}$ mJy Hz arcmin$^{-2}$) 
and $S_{1420} \geq$ 10 mJy arcmin$^{-2}$. The derived extinctions range from 
$A_{\rm V} \sim$ 1.0 to $A_{\rm V}$ = 2.8 mag. The {\it average} extinction 
$A_{\rm V mean} = 2.0$ mag is in excellent agreement with the mean values
of 2 mag apparent from the low-resolution (11$'$) extinction map 
published by Dickel $\&$ Wendker (1978). Dust emission at $\lambda 100 \mu$m
is also shown Fig. \ref{extinction}; the similarity between the two
images suggest that most if not all of the depicted dust is in front
of Simeis~57. At the same time, however, the similarity of the {\it extinction 
map} to the radio map also suggests that much of the material in front
of the nebula is actually closely associated with it.  

A more quantitative comparison of foreground extinction and 100$\mu$m 
dust emission follow from Fig. \ref{extinction} where we have plotted
$A_{\rm V}$ as a function of the far-infrared surface brightness at 
100 $\mu$m, $\sigma_{100}$.
We found a reasonably good correlation from which an average ratio 
$A_{\rm V}/\sigma_{100}$ = 0.016 mag/(MJy/ster) can be determined. 
About 1.5 mag of extinction is not directly related to Simeis~57,
but should probably be ascribed to cold, extended foreground dust.
We will discuss the extinction and far-infrared properties of 
Simeis~57 in more detail in a subsequent paper.

\section{Conclusions}

\begin{enumerate}
\item We have obtained high-resolution maps of the Galactic nebula 
Simeis~57 (= HS~191) at various radio continuum frequencies between 408 
and 1420 MHz. Analysis of these data, and those gleaned from existing
databases, shows the emission observed from the peculiar and complex
nebula to be wholly thermal.
\item Radio and optical images, including those taken in the
H$\alpha$ line of excited hydrogen, are very similar.
\item Although neither distance nor source of excitation of Simeis~57
are known, electron densities and masses seem to be moderate, of the 
order of 100 $\cc$ and a few solar masses respectively. The extended
emission may represent a gas mass up to a few dozen solar masses.
Emission measures do not exceed 5000 pc cm$^{-6}$.
\item In front of Simeis~57, extinction varies from $A_{\rm V}$ = 1.0
mag to $A_{\rm V}$ = 2.8 mag with a mean of about 2 mag. Extinction and
far-infrared ($\lambda 100\mu$m) emission are well-correlated. Although
much of the dust appears to be in front of the nebula, it is nevertheless
closely associated with it.
\end{enumerate}

\acknowledgements
Roeland Rengelink kindly assisted us in extracting the 327 MHz map from
the WENSS database. We are indebted to John Simonetti for permission to
reproduce the Virginia Tech Spectral-Line Survey (VTSS) image; the VTSS 
is supported by the National Science Foundation (see 
http://www.phys.vt.edu/\verb1~1halpha/)


\begin{thebibliography}{} 
%
\bibitem{} Cohen R.S., Dame T.M., Garay G., et al. 1988 \apjl 331, L95
\bibitem{} Cornwell T.J., 1988 \aua 202, 316
\bibitem{} Davies R.D., Elliott H.K., $\&$ Meaburn J., 1976 \mnras 81, 89
\bibitem{} Dennision B., Simonetti, J.H. $\&$ Topasna G., 1998, \pasa 15, 147 
\bibitem{} Dickel H.R. $\&$ Wendker H.J., 1977 \auas 29, 209
\bibitem{} Dickel H.R. $\&$ Wendker H.J., 1978 \aua 66, 289
\bibitem{} Dickel H.R., Wendker H. $\&$ Bieritz J.H., 1969 \aua 1, 270
\bibitem{} Dixon K. L., Johnson P. G. $\&$ Songsathaporn, R. 1981 
           Ap$\&$SS 78, 189 
\bibitem{} Gaze V. F. $\&$ Shajn G. A. 1951 Izv. Krym. Astrofiz. Obs. 7, 93
\bibitem{} Gaze V. F. $\&$ Shajn G. A. 1955 Izv. Krym. Astrofiz. Obs. 15, 11
\bibitem{} Goudis C., 1976 Ap$\&$SS 44, 281
\bibitem{} Habing H. J. $\&$ Israel F. P., 1979 \araa 17, 345
\bibitem{} Higgs. L.A., 1989 J. Roy. Astron. Soc. Can. 83, 105
\bibitem{} Higgs. L.A., Landecker T.L., Israel F.P. $\&$ Bally J. 1991
           J. Roy. Astron. Soc. Can. 85, 24
\bibitem{} Karyagina S.V. $\&$ Glushkov Yu. I., 1971 Astron. Tsirk. 
           No. 632, p. 1
\bibitem{} Landecker T.L., Dewdney P.E., Burgess T.A., et al. 2000 \auas 
           145, 509
\bibitem{} Langston G. I., Minter A. H. $\&$ Freismuth T. M. 2000 \aj 119, 2801
\bibitem{} Mezger P. G. $\&$ Henderson A.P., 1967 \apj 147, 471
\bibitem{} Pengelly R.M., 1964 \mnras 127, 145
\bibitem{} Piepenbrink A. $\&$ Wendker H.J., 1988 \aua 191, 313
\bibitem{} Rengelink R.B., Tang Y., de Bruyn A.G., Miley G.K., Bremer M.N.,
           R\"ottgering H.J.A. $\&$ Bremer M.A.R. 1997 \auas 124, 259
\bibitem{} Wendker H., 1967 Zschft f\"ur Ap. 66, 379
\bibitem{} Wendker H.J., 1970 \aua 4, 378
\bibitem{} Wendker H.J., Higss L.A. $\&$ Landecker T.L. 1991 \aua 241, 551
\bibitem{} Westerhout G., 1958 Bull. Astron. Inst. Neth. 14, 215
\end{thebibliography}
\end{document}